\documentclass[preprint,amsmath,showpacs]{revtex4}
\usepackage{graphicx}
\usepackage{dcolumn}
\usepackage{bm}

\def\bee{\begin{eqnarray}}
\def\eee{\end{eqnarray}}

\begin{document}
\draft
\title{The matter effects to neutrino oscillations $\nu_{\mu}\to \nu_e, \nu_{\mu}$ at
very long baselines and the neutrino mixing angles $\theta_{13}$ and
$\theta_{23}$ }
\author{Guey-Lin Lin\footnote{E-mail: glin@cc.nctu.edu.tw}
and Yoshiaki Umeda\footnote{E-mail: umeda@faculty.nctu.edu.tw} }
\address{
Institute of Physics, National Chiao-Tung University, Hsinchu 300,
Taiwan}
\date{\today}
\begin{abstract}
We study the matter effects to neutrino oscillations $\nu_{\mu}\to
\nu_{e}$ and $\nu_{\mu}\to \nu_{\mu}$ at very long baselines. It has
been observed that, for a very long baseline $L\simeq 10000$ km, the
resonance peak for the $\nu_{\mu}\to \nu_{e}$ oscillation and the
local minimum of the $\nu_{\mu}$ survival probability occurs at
similar energies. With a good knowledge on the absolute value of
$\Delta m_{31}^2$, measurements of the above oscillation
probabilities, $P_{\mu e}$ and $P_{\mu\mu}$, can be performed with
the neutrino energy tuned to the resonance peak of the $\nu_{\mu}\to
\nu_{e}$ oscillations. We show that the variations of CP violating
phase and the solar neutrino mixing parameters have negligible
effects on $P_{\mu e}$ and $P_{\mu\mu}$ for such a long baseline.
Hence $P_{\mu e}$ and $P_{\mu\mu}$ together determine the mixing
angles $\theta_{13}$ and $\theta_{23}$. Around the resonance peak
for $\nu_{\mu}\to \nu_e$ oscillations, the dependencies of $P_{\mu
e}$ and $P_{\mu\mu}$ on the above mixing angles are worked out in
detail, and the implications of such dependencies are discussed.
\end{abstract}
\maketitle
The understanding of neutrino masses and mixing matrix is crucial to
unveil the mystery of lepton flavor structure. The updated SKK
analysis of the atmospheric neutrino data gives \cite{Ashie:2004mr}
\begin{equation}
 1.9\cdot 10^{-3}\,\,\, {\rm eV}^{2}< \vert \Delta m_{31}^{2}\vert < 3.0\cdot
 10^{-3}\,\,\, {\rm eV}^{2},  \ \sin^{2}2\theta_{23} >0.9. \label{range23}
\end{equation}
This is a  $90\% \, {\rm C.L.}$ range with the best fit values given
by $\sin^{2}2\theta_{23} =1$ and $\Delta m_{31}^{2}=2.4\cdot
10^{-3}\, \, \, {\rm eV}^{2}$ respectively. The scenario of
$\nu_{\mu}\to \nu_{\tau}$ oscillation for atmospheric neutrinos has
been confirmed by the K2K experiment \cite{Aliu:2004sq}. Furthermore
the results in the solar neutrino oscillation measurements are also
confirmed by KamLAND reactor measurements
\cite{Eguchi:2002dm,Araki:2004mb}. Combining these measurements, the
LMA solution of the solar neutrino problem is established and the
updated $3\sigma$ parameter ranges are given by
\cite{Bahcall:2004ut}
\begin{equation}
7.4\cdot 10^{-5}\,\,\, {\rm eV}^{2}< \Delta m_{21}^{2}< 9.2\cdot
10^{-5}\, \, \, {\rm eV}^{2}, \ 0.28< \tan^{2}\theta_{12}< 0.58,
\label{range12}
\end{equation}
with the best fit values $\Delta m_{21}^{2}=8.2\cdot 10^{-5}\, \, \,
{\rm eV}^{2}$ and $\tan^{2}\theta_{12}=0.39$.

Despite the achievements so far in measuring the neutrino mixing
parameters, the sign of $\Delta m_{31}^2$, the mixing angle
$\theta_{13}$ and the CP violating parameter $\delta_{\rm CP}$ in
the mixing matrix remain to be determined. The sign of $\Delta
m_{31}^2$ can be determined through matter effects to $\nu_{\mu}\to
\nu_e$ ($\bar{\nu}_{\mu}\to \bar{\nu}_e$) and $\nu_e\to \nu_{\mu}$
($\bar{\nu}_e\to \bar{\nu}_{\mu}$) oscillations in the very long
baseline experiments \cite{Mocioiu:2000st,Freund:1999gy}. However,
such oscillations are sensitive to the mixing angle $\theta_{13}$,
which has been constrained by the reactor experiments
\cite{Apollonio:1999ae,Boehm:2001ik}. The CHOOZ experiment
\cite{Apollonio:1999ae} gives a more stringent constraint on
$\theta_{13}$ with $\sin^2 2\theta_{13} < 0.1$ for a large $\Delta
m_{31}^2$ (90\% C.L.). A recent global fit gives $\sin^2
2\theta_{13}< 0.09 \ (0.18)$ at $90\%$ C.L. ($3\sigma$)
\cite{Maltoni:2004ei}.

It is known that the oscillation $\nu_{\mu}\to \nu_e$, for example,
is enhanced (suppressed) by the matter effect for $\Delta m_{31}^2
>0$ ($\Delta m_{31}^2 <0$). On the contrary, the $\bar{\nu}_{\mu}\to
\bar{\nu}_e$ oscillation is enhanced (suppressed) by the matter
effect for $\Delta m_{31}^2 <0$ ($\Delta m_{31}^2 >0$). In the
following discussions we shall assume $\Delta m_{31}^2 >0$ unless
explicitly specified. It is noteworthy that, although the matter
enhanced angle $\theta_{13}^m$ reaches to $\pi/4$ at the resonance
energy, the oscillation probability, $P(\nu_{\mu}\to \nu_e)\equiv
P_{\mu e}$, remains to be small unless for a very long baseline.
Indeed, it is demonstrated that \cite{Banuls:2001zn}, in the
constant density approximation for the Earth density profile, the
condition for a maximal $P_{\mu e}$ reads:
\begin{equation}
\rho L^{\rm max}\simeq \frac{(2n+1)\pi}{\tan2\theta_{13}}\times
5.18\cdot 10^3 \ {\rm km}\cdot {\rm g/cm}^3,
\end{equation}
where $\rho$ is the averaged mass density of the Earth, $L^{\rm
max}$ is the required oscillation length, and $n$ is an integer. For
$n=0$ and $\sin^2 2\theta_{13}=0.1$, $L^{\rm max}$ is found to be
$10200$ km \cite{Gandhi:2004md}. A smaller $\sin^2 2\theta_{13}$
requires an even larger oscillation length. Clearly, to take a full
advantage of the matter enhancement, one is led to consider
experiments with a baseline of the order of the Earth diameter
\cite{DeJongh:2002dv}.

It has been demonstrated that the matter effects to $P(\nu_{\mu}\to
\nu_{\tau})\equiv P_{\mu\tau}$ and $P(\nu_{\mu}\to \nu_{\mu})\equiv
P_{\mu\mu}$ can also be significant in the vicinity of $P_{\mu e}$'s
resonance peak \cite{Gandhi:2004md,Gandhi:2004bj}. In fact, the
criterion for the maximal matter effect to $P_{\mu\tau}$ is
identified to be $E_{\rm res}\simeq E_{\rm peak}^{\rm vac}$ where
the former is the resonance energy for the oscillation $P_{\mu e}$
while the latter is the energy that the vacuum approximation of
$P_{\mu\tau}$ attains its peak value. Such a criterion amounts to
\cite{Gandhi:2004md,Gandhi:2004bj}
\begin{equation}
\rho L^{\rm max}\simeq (2n+1)\pi (\cos2\theta_{13})\times 5.18\cdot
10^3 \ {\rm km}\cdot {\rm g/cm}^3.
\end{equation}
For $n=0$ and $\sin^2 2\theta_{13}=0.1$, $L^{\rm max}\simeq 4400$
km, while, for $n=1$ with the same $\sin^2 2\theta_{13}$ value,
$L^{\rm max}\approx 9700$ km. In the latter case, $P_{\mu\tau}$ is
decreased from its vacuum oscillation value by as much as $70\%$.
This implies that the muon neutrino survival probability
$P_{\mu\mu}$ also receives significant matter effects in the same
energy range.

It is interesting to note that the matter effect to $P_{\mu e}$ and
that to $P_{\mu\mu}$ are both significant in the same energy range
for baselines of the order $10^4$ km. Hence simultaneous
measurements of $P_{\mu e}$ and $P_{\mu\mu}$ in such a baseline
provide a stringent test on the matter effect, which is the focus of
this work. To discuss these oscillations, we begin with the relation
connecting flavor and mass eigenstates of neutrinos,
$\nu_{\alpha}=\sum_i U_{\alpha i}\nu_{i}$, with $U$ the
Pontecorvo-Maki-Nakagawa-Sakata mixing matrix given by
\begin{equation}
U =\left(\!
\begin{array}{ccc}
c_{12}c_{13} & s_{12}c_{13} & s_{13}e^{-i\delta_{\rm CP}} \\
-s_{12}c_{23}-c_{12}s_{13}s_{23}e^{i\delta_{\rm CP}} &
c_{12}c_{23}-s_{12}s_{13}s_{23}e^{i\delta_{\rm CP}}
& c_{13}s_{23} \\
s_{12}s_{23}-c_{12}s_{13}c_{23}e^{i\delta_{\rm CP}} &
-c_{12}s_{23}-s_{12}s_{13}c_{23}e^{i\delta_{\rm CP}} & c_{13}c_{23}
\end{array}
\!\right)\,,
\end{equation}
where $s_{ij}$ and $c_{ij}$ denote $\sin\theta_{ij}$ and
$\cos\theta_{ij}$, respectively. For the Dirac type CP-phase
$\delta_{\rm CP}$, we allow its value ranging from 0 to 2$\pi$. The
evolutions of neutrino flavor eigenstates are governed by the
equation
\begin{eqnarray}
i\frac{d}{dt} |\nu(t)\rangle &=&  \left\{ \frac{1}{2E_\nu}U \left(\!
\begin{array}{ccc}
0 & 0               & 0 \\
0 & \Delta m^2_{21} & 0 \\
0 & 0               & \Delta m^2_{31}
\end{array}
\right)U^\dagger  + \left(\!
\begin{array}{ccc}
V & 0 & 0 \\
0 & 0 & 0 \\
0 & 0 & 0
\end{array}
\!\right)\right\} |\nu(t)\rangle \,, \label{evolution}
\end{eqnarray}
where $|\nu(t)\rangle = (\nu_e(t), \nu_\mu(t), \nu_\tau(t))^T$,
$\Delta m^2_{ij} \equiv m^2_{i}-m^2_{j}$ is the mass-squared
difference between the $i$-th and $j$-th mass eigenstates, and
$V\equiv \sqrt{2}G_F N_e$ is the effective potential arising from
the charged current interaction between $\nu_e$ and electrons in the
medium with $N_e$ the electron number density. Numerically $V= 7.56
\times 10^{-14}$ $(\rho/[{\rm g/cm^3}])$$Y_e$[{\rm eV}]  with $Y_e$
denoting the number of electrons per nucleon. For the Earth matter,
$Y_e\sim 0.5$. One solves Eq.~(\ref{evolution}) by diagonalizing the
Hamiltonian on its right hand side. This amounts to writing the
right hand side of Eq.~(\ref{evolution}) as
$U'H'U^{'\dagger}|\nu(t)\rangle$ with $U'$ the neutrino mixing
matrix in the matter and $H' \equiv {\rm diag}(E_1, E_2, E_3)$ the
Hamiltonian after diagonalization. To obtain various oscillation
probabilities described later, we have used the parametrization in
\cite{Gandhi:1995tf} for the Earth density profile \cite{profile}.
For illustration, we shall take $L=9300$ km, which is a distance
between Fermilab and Kamioka \cite{DeJongh:2002dv}. This distance is
in fact very ideal for studying $P_{\mu e}$ and $P_{\mu\mu}$ since
both oscillations receive significant matter effects around such a
distance.

\begin{figure}[t]
\begin{center}
$\begin{array}{c}
\includegraphics*[width=9.5cm]{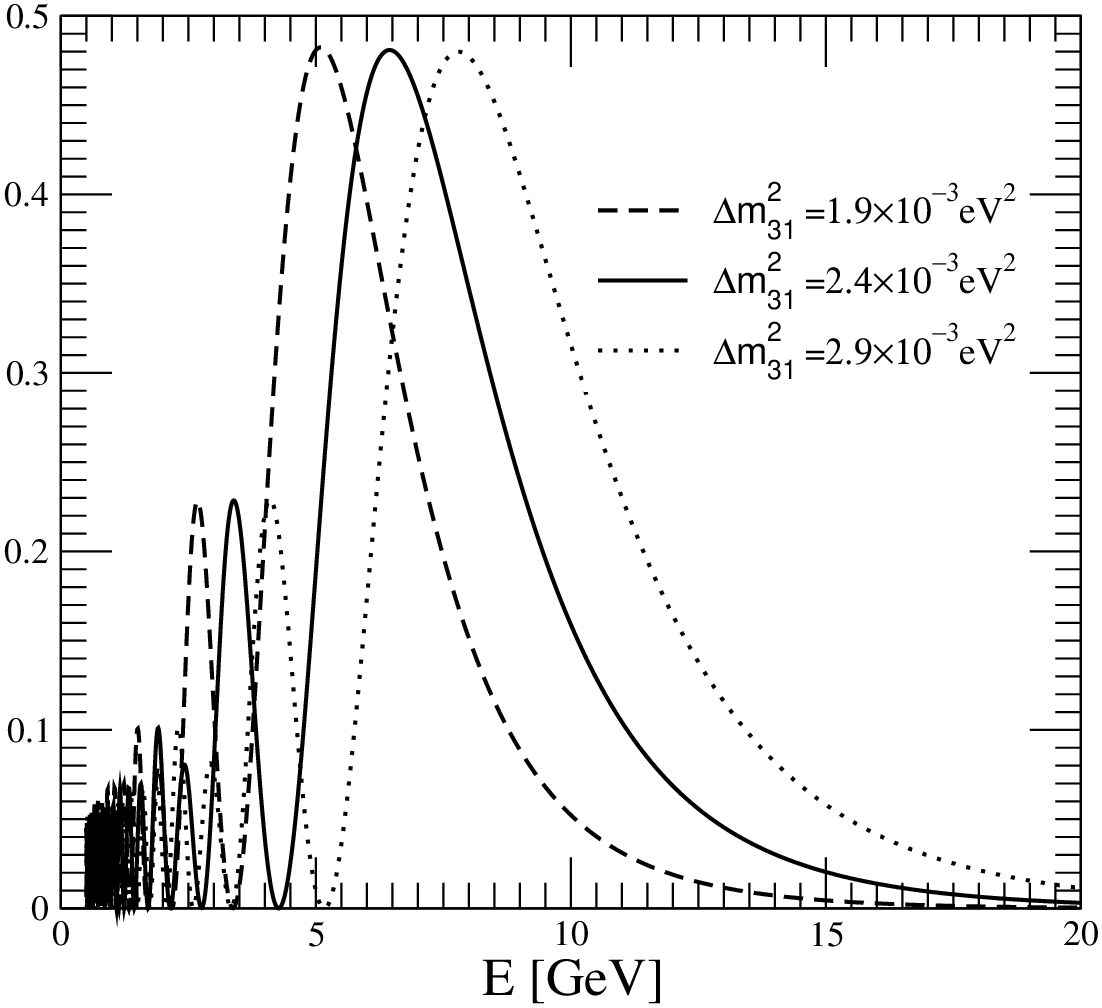}
\end{array}$
\end{center}
\caption{The $\Delta m^2_{31}$ dependence of the transition
probability $P_{\mu e}$ for $\sin^2 2\theta_{13}=0.1$, $\sin^2
2\theta_{23}=1$, $\delta_{\rm CP}=0$, and the baseline length
$L=9300$ km. }\label{fig:m31}
\end{figure}

To measure $P_{\mu e}$ and $P_{\mu\mu}$ near the resonance peak of
the $\nu_{\mu}\to \nu_e$ oscillation, it is crucial to know the
precise value of $\Delta m^2_{31}$. The current range for this
parameter, Eq.~(\ref{range23}), results in a corresponding range for
the peak energy of $P_{\mu e}$ as shown in Fig.~\ref{fig:m31}.
The energy value at the peak of $P_{\mu e}$ is proportional to
$\Delta m^2_{31}$ as dictated by the resonance condition
$2E_{\nu}V=\Delta m_{31}^2\cos\theta_{13}$ while the maximal
magnitude of $P_{\mu e}$ is independent of $\Delta m^2_{31}$. This
is easily understood by inspecting the expression for $P_{\mu e}$ in
the constant density approximation \cite{Petcov:1987cd}:
\begin{equation}
P_{\mu e}=\sin^2\theta_{23}\sin^2 2\theta_{13}^{\rm
m}\sin^2\left(1.27\Delta_{31}^{\rm m}L/E_{\nu}\right), \label{pmue}
\end{equation}
with $L$ in km, $E_{\nu}$ in eV,
\begin{equation}
\Delta_{31}^{\rm m}=\sqrt{\left(\Delta m_{31}^2\cos
2\theta_{13}-2E_{\nu}V\right)^2+\left(\Delta m_{31}^2\sin
2\theta_{13}\right)^2},
\end{equation}
and
\begin{equation}
\sin 2\theta_{13}^{\rm m}=\frac{\sin 2\theta_{13}\cdot \Delta
m_{31}^2}{\Delta_{31}^{\rm m}}.
\end{equation}
Clearly, at the resonance energy, the last factor on the right hand
side of Eq.~(\ref{pmue}) does not depend on $\Delta m_{31}^2$.

\begin{figure}[tbp]
\begin{center}
$\begin{array}{c}
\includegraphics*[width=9.5cm]{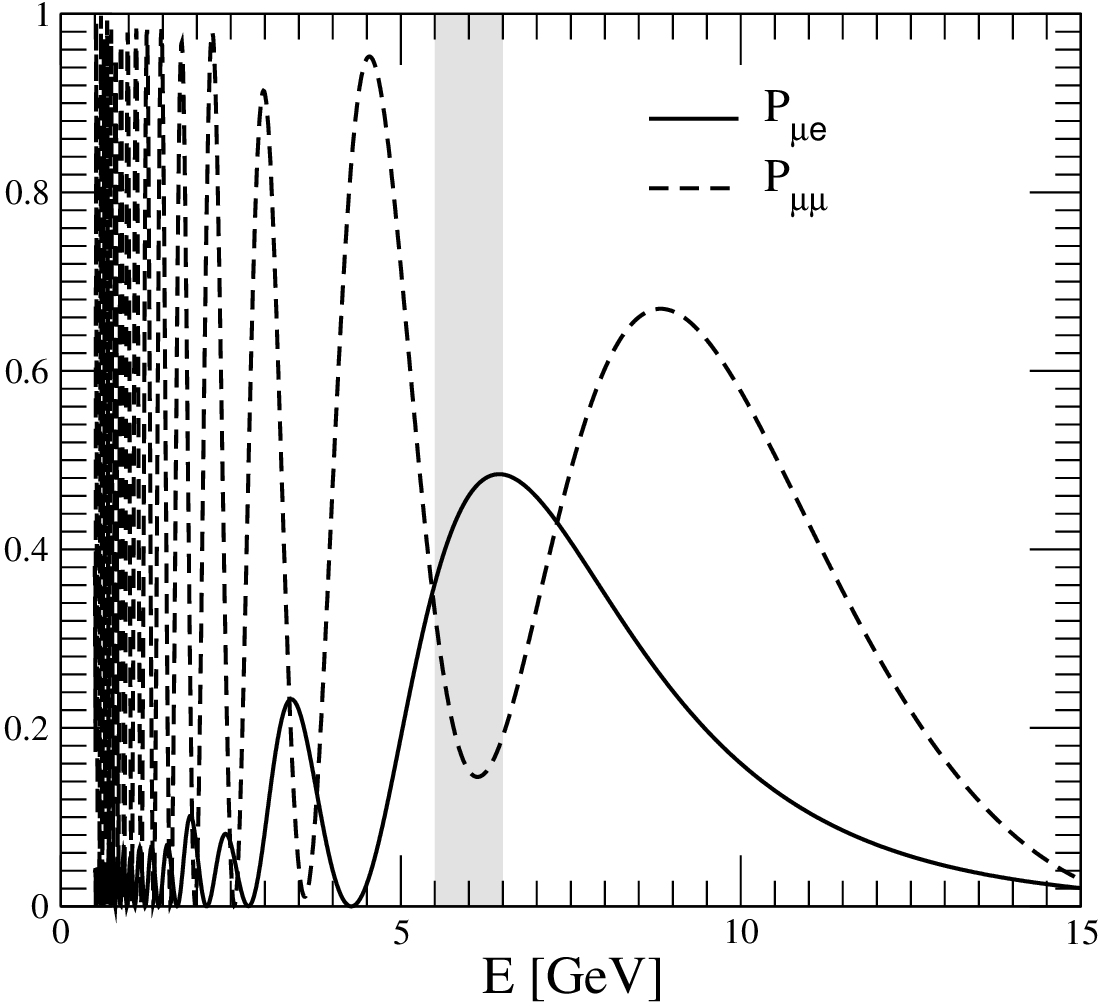}
\end{array}$
\end{center}
\caption{The transition probabilities $P_{\mu e}$ and $P_{\mu \mu}$
for a baseline length $L=9300$ km with $\sin^2 2\theta_{13}=0.1$,
$\sin^2 2\theta_{23}=1$ and $\delta_{\rm cp}=0$. The gray zone
represents the region with the neutrino energy $E_{\nu}$ ranging
from $5.5$ GeV to $6.5$ GeV, where the maximum of $P_{\mu e}$ and
the local minimum of $P_{\mu\mu}$ are located.} \label{integ}
\end{figure}

As pointed out earlier, the matter effects to $P_{\mu e}$ and
$P_{\mu\mu}$ are both significant in the same energy range for our
interested baseline length. This is indicated by the gray zone of
Fig.~\ref{integ} where the energy dependencies of $P_{\mu e}$ and
$P_{\mu\mu}$ are depicted.
The probability $P_{\mu\mu}$ shows an intriguing feature in this
energy zone. Due to the large matter effect, $P_{\mu\mu}$ does not
drop to zero while reaching to its local minimum. For $\Delta
m_{31}^2=2.4\times 10^{-3}$ eV$^2$, $\sin^2 2\theta_{13}=0.1$ and
$\sin^2 2\theta_{23}=1$, the probability $P_{\mu e}$ peaks at $6.4$
GeV while the local minimum for $P_{\mu\mu}$ is at $6.1$ GeV. We
have varied the values for $\theta_{13}$ and $\theta_{23}$. The
energy value for the maximal $P_{\mu e}$ and that for the local
minimum of $P_{\mu\mu}$ are shifted. However, the shifts are within
$300$ MeV for both cases.

\begin{figure}[tbp]
\begin{center}
$\begin{array}{c}
\includegraphics*[width=9.5cm]{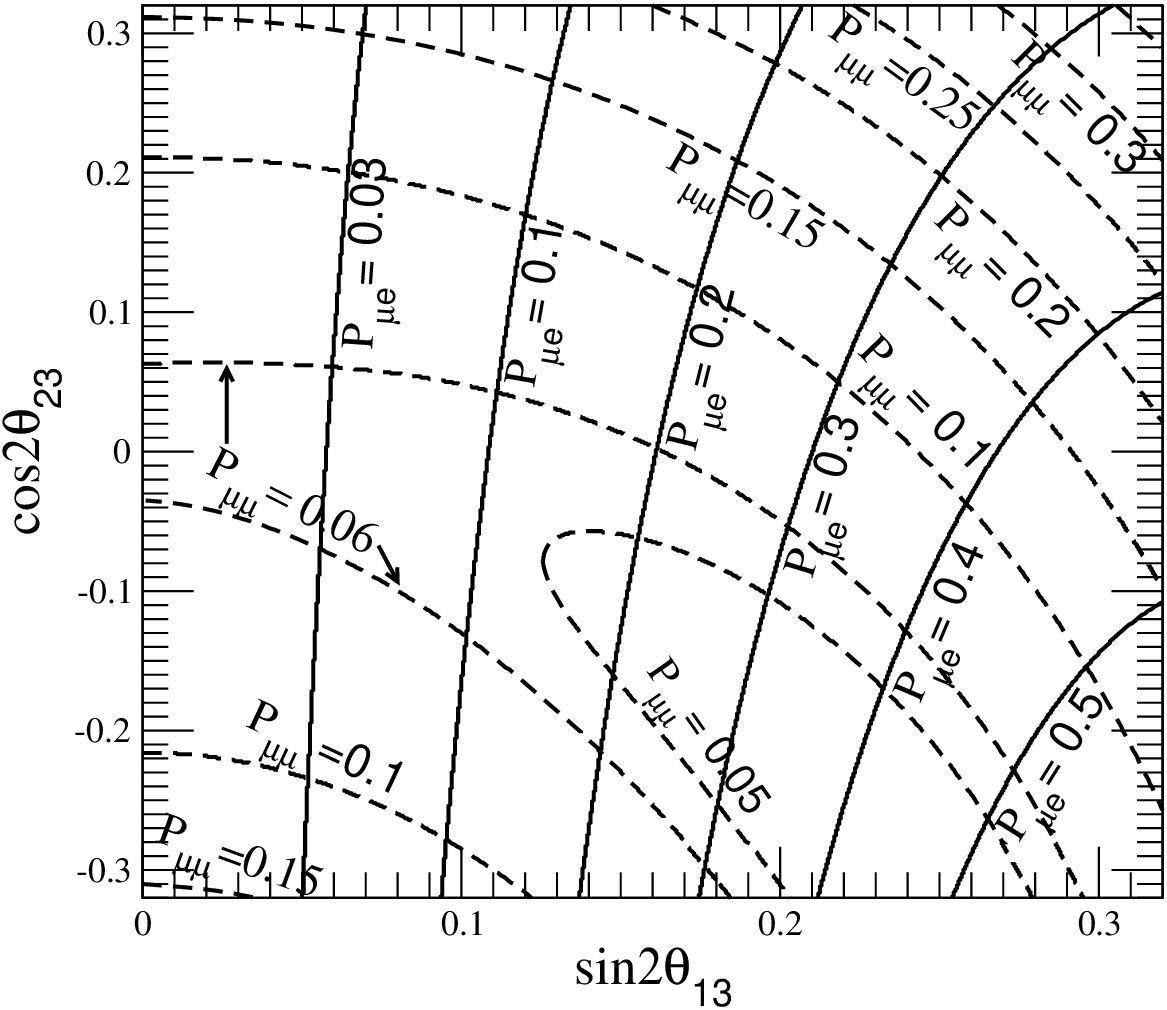}
\end{array}$
\end{center}
\caption{The contour graph of the transition probabilities $P_{\mu
e}$(solid line) and $P_{\mu \mu}$(dashed line) in the $\sin
2\theta_{13}$-$\cos 2\theta_{23}$ plane for $\delta_{\rm CP}=0$ and
the baseline length $L=9300$ km. The plotted $P_{\mu e}$ and $P_{\mu
\mu}$ are the averaged probabilities for neutrino energies ranging
from $5.5$ GeV to $6.5$ GeV }\label{fig:contour}
\end{figure}

The probabilities $P_{\mu\mu}$ and $P_{\mu e}$ as functions of
$\theta_{23}$ and $\theta_{13}$ are shown in Fig.~\ref{fig:contour}.
The range for $\sin 2\theta_{13}$ is taken from $0$ to $0.32$ in
accordance with the reactor bound $\sin^2 2\theta_{13}< 0.1$
\cite{Apollonio:1999ae}, while that for $\cos 2\theta_{23}$ is from
$-0.32$ to $0.32$ which is equivalent to $\sin^2 2\theta_{23}> 0.9$
\cite{Ashie:2004mr}.
The oscillation probabilities plotted here are the averaged ones for
neutrino energies ranging from $5.5$ GeV to $6.5$ GeV. For $\Delta
m_{31}^2=2.4\times 10^{-3}$ eV$^2$, both $P_{\mu e}$ and
$P_{\mu\mu}$ receive large matter effects. We argue that these
probabilities do reflect the relative event rates. In fact, to
obtain neutrino event rates, one has to convolve the oscillation
probabilities with the initial muon neutrino flux and the conversion
rates from neutrinos to charged leptons. However, since our
interested energy range is narrow enough, the neutrino-nucleon
scattering cross section can be treated as a constant. The same is
true for the $\nu_{\mu}$ flux. We note that $P_{\mu e}$ is
essentially only sensitive to $\sin 2\theta_{13}$ for $P_{\mu e}\leq
0.1$.
\begin{figure}[tbp]
\begin{center}
$\begin{array}{c}
\includegraphics*[width=9.5cm]{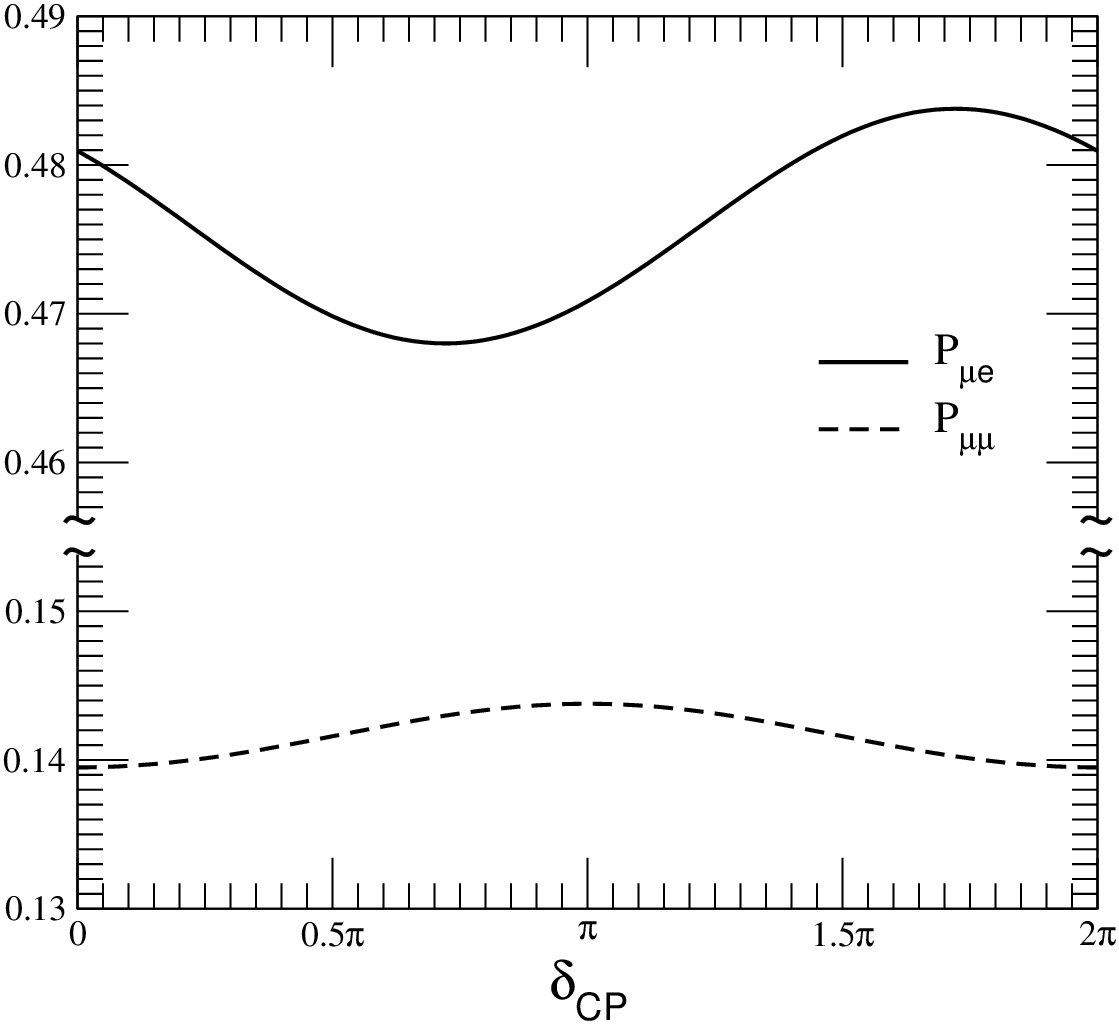}
\end{array}$
\end{center}
\caption{The CP-phase dependencies of the transition probabilities
$P_{\mu e}$ and $P_{\mu \mu}$ for $\sin^2 2\theta_{13}=0.1$, $\sin^2
2\theta_{23}=1$, and the baseline length $L=9300$ km.}
\label{fig:cp}
\end{figure}
However $P_{\mu e}$ is also sensitive to $\cos 2\theta_{23}$ as it
gets larger. The contours for constant $P_{\mu\mu}$'s show more
complicated structures. We observe that, for a fixed $\theta_{13}$,
a given $P_{\mu\mu}$ yields two solutions for $\theta_{23}$ for
$P_{\mu\mu}\leq 0.15$. The measurement of $P_{\mu e}$ might help to
remove this degeneracy if $P_{\mu e}$ is large and consequently
sensitive to $\cos 2\theta_{23}$. We also note that the above
degeneracy no longer exists for $P_{\mu\mu}> 0.15$.
\begin{figure}[tbp]
\begin{center}
$\begin{array}{c}
\includegraphics*[width=9.5cm]{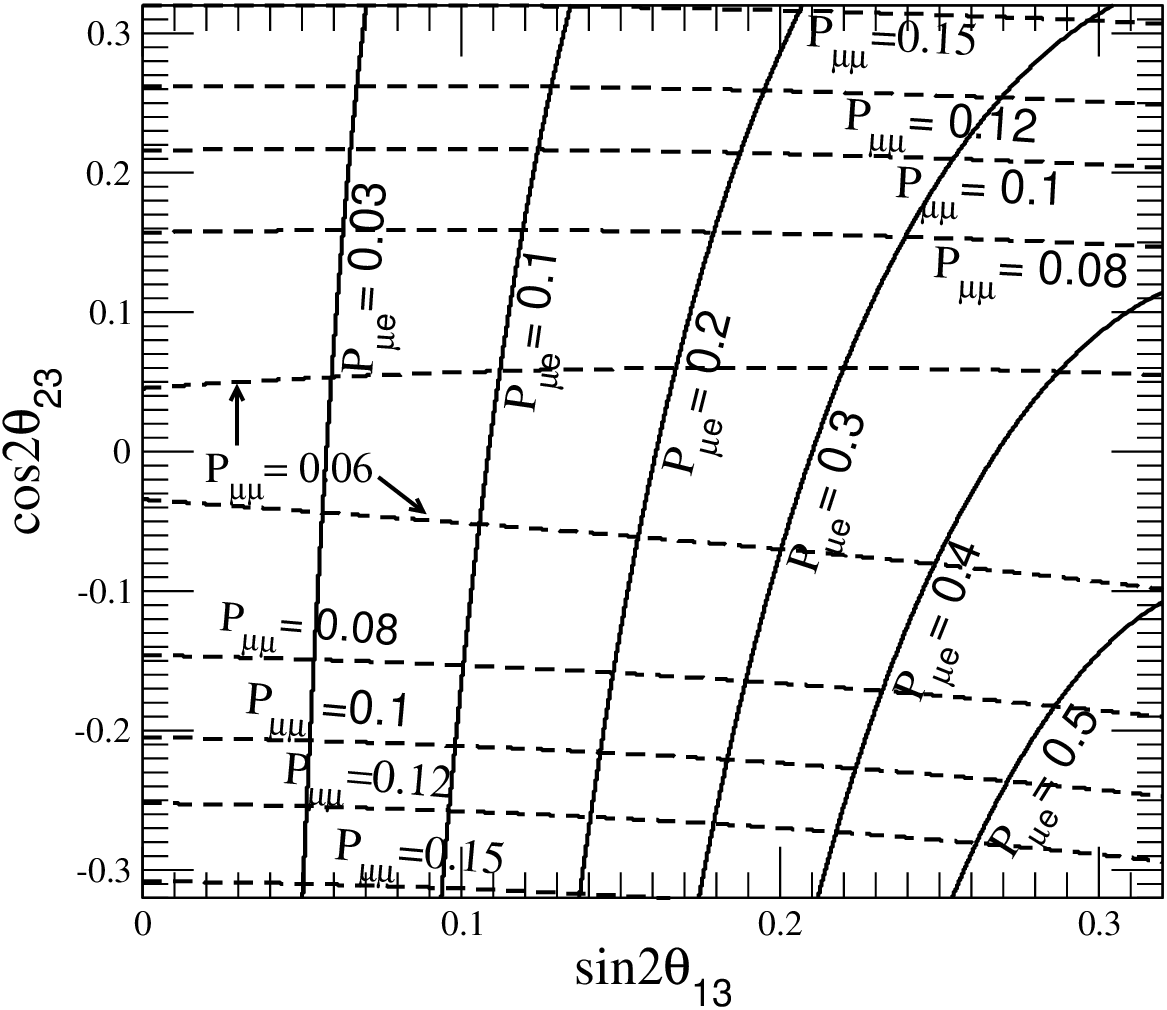}
\end{array}$
\end{center}
\caption{The contour graph of the transition probabilities $P_{\mu
e}$(solid line) and $P_{\mu \mu}$(dashed line) for $\delta_{\rm
CP}=0$ and the baseline length $L=9300$ km. Vacuum oscillation is
assumed for $P_{\mu\mu}$.}\label{fig:vacuum}
\end{figure}

We point out that simultaneous measurements of $P_{\mu e}$ and
$P_{\mu\mu}$ determine $\theta_{13}$ and $\theta_{23}$. It is found
that the variations of $\Delta m_{21}^2$ and $\theta_{12}$ give
negligible effects on the above oscillation probabilities.  The
effects of CP violating phase $\delta_{\rm CP}$ on $P_{\mu e}$ and
$P_{\mu\mu}$ are also studied. With $\Delta m_{21}^2=8.2\cdot
10^{-5}\, \, \, {\rm eV}^{2}$, $\sin^2 2\theta_{23}=1$, $\sin^2
2\theta_{13}=0.1$, the dependencies of the peak value of $P_{\mu e}$
and the corresponding minimal value of $P_{\mu\mu}$ on $\delta_{\rm
CP}$ are shown in Fig.~\ref{fig:cp}. It is seen that the peak value
of $P_{\mu e}$ differs by no more than $4\%$ for the entire range of
$\delta_{\rm CP}$. Similarly, the corresponding minimal value of
$P_{\mu\mu}$ differs by no more than $3\%$.

The matter effect to $P_{\mu e}$ can be identified by the resonance
enhancement to the oscillation $\nu_{\mu}\to \nu_e$, given our
assumption that $\Delta m_{31}^2 > 0$. On the other hand, the matter
effect to $P_{\mu\mu}$ is seen by comparing Fig.~\ref{fig:contour}
with a corresponding plot, Fig.~\ref{fig:vacuum}, where matter
effects to $P_{\mu\mu}$ are neglected. In Fig.~\ref{fig:vacuum}, one
can see that $P_{\mu\mu}$ is  not sensitive to $\sin 2\theta_{13}$
and the contours for constant $P_{\mu\mu}$'s behave very differently
from those in Fig.~\ref{fig:contour}, particularly for a larger
$\sin2\theta_{13}$. We observe that, in the vacuum oscillation case,
$P_{\mu\mu}\leq 0.15$ for almost all the parameter space in the
$\sin 2\theta_{13}$-$\cos 2\theta_{23}$ plane. Hence an observation
of $P_{\mu\mu}$ significantly greater than $0.15$ is a signature of
the matter effect. The parameter space for this to occur is situated
in the upper right corner of $\sin 2\theta_{13}$-$\cos 2\theta_{23}$
plane. This is a region with $\theta_{23}< \pi/4$.

The matter effect to $P_{\mu\mu}$ can also be tested in other
parameter regions. This is true when values of $\theta_{13}$ and
$\theta_{23}$ are accurately measured in other future experiments.
There are extensive discussions on measuring $\theta_{13}$ in
reactor neutrino experiments \cite{Anderson:2004pk,Oberauer:2005hr}.
The issue of determining $\theta_{23}$ in the three-flavor
oscillation framework has also been discussed in
\cite{Huber:2004ug,Antusch:2004yx,Minakata:2004pg}. It is expected
that $\sin^2 2\theta_{23}$ can be measured to an $1\%$ accuracy in
the J-PARC$\to$ Super-Kamiokande (JPARC-SK) experiment
\cite{Itow:2001ee}. This translates into a $10\%$ accuracy for the
determination of $\sin^2 \theta_{23}$ for $\theta_{23}$ around
$\pi/4$ at $90\%$ C.L., provided $\sin^2 2\theta_{13}$ is accurately
measured by the reactor experiment \cite{Minakata:2004pg}, say
around $2\%$ at $90\%$ C.L. according to the analysis in
\cite{Minakata:2003wq}. Given the future accuracies for determining
$\sin^2 \theta_{23}$ and $\sin^2 2\theta_{13}$, it is possible to
identify the matter effect in the muon survival probability
$P_{\mu\mu}$. For illustration, let us assume, for example, $P_{\mu
e}$ and $P_{\mu\mu}$ are measured to be $0.4$ and $0.1$ respectively
in the very long baseline experiment, with some uncertainties
understood for each measurement. Taking into account the matter
effect to $P_{\mu\mu}$, these oscillation probabilities correspond
to $(\sin 2\theta_{13}, \cos 2\theta_{23})=(0.26,-0.03)$ according
to Fig.~\ref{fig:contour}. However, by assuming vacuum oscillations
for $P_{\mu\mu}$, one obtains $(\sin 2\theta_{13}, \cos
2\theta_{23})=(0.22,-0.22)$ according to Fig.~\ref{fig:vacuum}.
Therefore the extracted central values of $\sin 2\theta_{13}$ differ
by $16\%$ with and without the matter effects to $P_{\mu\mu}$ taken
into account. This discrepancy is considerably larger than the
expected uncertainty in the determination of $\sin 2\theta_{13}$ by
the reactor experiment, which is approximately $4\%$ with the
above-mentioned uncertainty $\delta(\sin^2 2\theta_{13})\simeq 2\%$
and the central value $\sin 2\theta_{13}=0.26$. Hence the matter
effect to $P_{\mu\mu}$ can in principle be identified by the
extracted value of $\sin 2\theta_{13}$,  provided there are
sufficient accuracies in the measurements of $P_{\mu e}$ and
$P_{\mu\mu}$. Furthermore, we note that $\sin^2 \theta_{23}=0.52$
with the matter effect to $P_{\mu\mu}$ taken into account while
$\sin^2 \theta_{23}=0.61$ without the matter effect. The difference
on the extracted value of $\sin^2 \theta_{23}$ is about $17\%$,
which is also larger than the expected $10\%$ uncertainty in
determining $\sin^2 \theta_{23}$ by the JPARC-SK experiment.
Therefore, if one can accurately measure $P_{\mu e}$ and
$P_{\mu\mu}$, the matter effects to $P_{\mu\mu}$ can also be
identified through the extracted value of $\sin^2 \theta_{23}$.

In summary, we have discussed the matter effects to $\nu_{\mu}\to
\nu_{e}, \nu_{\mu}$ oscillations at very long baselines. We
illustrate these effects with a baseline $L=9300$ km, the distance
between Fermilab and Kamioka. Around this baseline length, the
matter effects to $P_{\mu e}$ and $P_{\mu\mu}$ are both significant
near the resonance energy for $\nu_{\mu}\to \nu_e$ oscillations. We
have worked out in detail the dependencies of these probabilities on
the mixing angles $\theta_{13}$ and $\theta_{23}$ with $\Delta
m_{31}^2$ taken to be $2.4\times 10^{-3}$ eV$^2$. We have also shown
that these probabilities are not sensitive to the variations of
other neutrino oscillation parameters. We have identified the
parameter space in the $\sin 2\theta_{13}$-$\cos 2\theta_{23}$ plane
in which the matter effect to $P_{\mu\mu}$ can be most easily
identified. We also argued that it is possible to identify the same
matter effect in the remaining parameter space provided
$\theta_{13}$ and $\theta_{23}$ are accurately measured in other
future neutrino experiments.

\section*{Acknowledgements}
We thank F. DeJongh for discussions. This work is supported by the
National Science Council of Taiwan under the grant numbers NSC
93-2112-M-009-001 and NSC 93-2811-M-009-017.

\end{document}